\documentclass[twocolumn,showpacs,preprintnumbers,amsmath,amssymb]{revtex4}

\usepackage{graphicx}
\usepackage{dcolumn}
\usepackage{bm}
\usepackage{type1cm}
\usepackage{bbm}

\DeclareMathOperator{\Tr}{Tr}

\DeclareMathOperator{\SU}{SU}

\DeclareMathOperator{\Real}{Re}
\DeclareMathOperator{\Imag}{Im}

\newcommand{\rmd}{{\rm d}}

\newcommand\fverb{\setbox\pippobox=\hbox\bgroup\verb}
\newcommand\fverbdo{\egroup\medskip\noindent%
                        \fbox{\unhbox\pippobox}\ }
\newcommand\fverbit{\egroup\item[\fbox{\unhbox\pippobox}]}
\newbox\pippobox

\begin{document}

\preprint{RIKEN-TH-121, OIQP-07-15}

\title{Euclidean lattice simulation for the dynamical supersymmetry breaking}%

\author{Issaku Kanamori}\email{kanamori-i@riken.jp}
\author{Hiroshi Suzuki}\email{hsuzuki@riken.jp}
\affiliation{%
Theoretical Physics Laboratory, RIKEN, Wako 2-1, Saitama 351-0198, Japan
}%

\author{Fumihiko Sugino}\email{fumihiko_sugino@pref.okayama.lg.jp}
\affiliation{%
Okayama Institute for Quantum Physics, Kyoyama 1-9-1, Okayama 700-0015, Japan
}%

\date{\today}

\begin{abstract}
The global supersymmetry is spontaneously broken if and only if the
ground-state energy is strictly positive. We propose to use this fact to
observe the spontaneous supersymmetry breaking in euclidean lattice
simulations. For lattice formulations that possess a manifest fermionic
symmetry, there exists a natural choice of a hamiltonian operator that is
consistent with a topological property of the Witten index. We confirm validity
of our idea in models of the supersymmetric quantum mechanics. We then examine
a possibility of a dynamical supersymmetry breaking in the two-dimensional
$\mathcal{N}=(2,2)$ super Yang-Mills theory with the gauge group~$\SU(2)$, for
which the Witten index is unknown. Differently from a recent conjectural claim,
our numerical result tempts us to conclude that supersymmetry is not
spontaneously broken in this system.
\end{abstract}

\maketitle

It is widely believed that supersymmetry is relevant in particle physics beyond
the standard model and it is spontaneously broken by some mechanism. If
supersymmetry is not spontaneously broken at the tree level of the loop
expansion, it remains so to all orders of the loop expansion. There still
exists, however, a possibility that supersymmetry is spontaneously broken
non-perturbatively. Precise study of such \emph{dynamical supersymmetry
breaking\/} remains elusive because we have no universal framework that defines
supersymmetric (especially gauge) theories at a non-perturbative level.

Generally, the Witten index $\Tr(-1)^F$~\cite{Witten:1982df}, where $F$ is the
fermion number operator, provides an important clue. One can infer that the
dynamical supersymmetry breaking does not occur in a wide class of
supersymmetric models for which the Witten index can be computed to be
non-zero. However, the Witten index is not a panacea. There exist physically
interesting models for which it is very difficult to determine the Witten index
and, in some cases, the index itself might be ill-defined due to a gap-less
continuous spectrum.

In this letter, we consider a possibility to observe the dynamical
supersymmetry breaking in euclidean lattice simulations, in the light of recent
developments on lattice formulation of supersymmetric
theories~\cite{Kaplan:2003uh,Feo:2004kx,Giedt:2006pd,Giedt:2007hz}. The
conceptually clearest way to observe the spontaneous supersymmetry breaking
would be to examine the degeneracy of bosons' and fermions' mass spectra in
two-point correlation functions. Here, we propose an alternative method that is
based on the following fact. The global supersymmetry is spontaneously broken
if and only if the ground-state energy is strictly
positive~\cite{Witten:1981nf}. In principle, therefore, one can judge whether
the supersymmetry breaking takes place or not if the ground-state energy can be
computed.

Let us first recall that the thermal average of the hamiltonian~$H$ with the
inverse temperature~$\beta$ is expressed by the euclidean functional integral
as~\footnote{The thermal average of the hamiltonian in $(1+0)$-dimensional
supersymmetric gauge theories has been numerically investigated in
Refs.~\cite{Catterall:2007fp,Anagnostopoulos:2007fw}.}
\begin{equation}
   \frac{\Tr He^{-\beta H}}{\Tr e^{-\beta H}}
   =\frac{\int_{\text{aPBC}}\rmd\mu\,He^{-S}}
   {\int_{\text{aPBC}}\rmd\mu\,e^{-S}}
   \equiv\langle H\rangle_{\text{aPBC}},
\label{one}
\end{equation}
where $S$ is the euclidean (i.e., imaginary-time) action and $\rmd\mu$
symbolically denotes a functional integral measure. In the right-hand side, the
time-period of the system is taken to be $\beta$. What is very important in
Eq.~(\ref{one}) is the boundary condition in the temporal direction. It must be
periodic for all bosonic variables and anti-periodic (aPBC) for all fermionic
variables. (For all bosonic variables and for all variables with respect to
spatial directions, we always assume the periodic boundary conditions.) In the
large imaginary-time or the low-temperature limit $\beta\to\infty$, only the
ground state(s) contributes to Eq.~(\ref{one}) and the ground-state
energy~$E_0$ is given by
\begin{equation}
   E_0=\lim_{\beta\to\infty}\langle H\rangle_{\text{aPBC}}.
\label{two}
\end{equation}
If $E_0>0$, supersymmetry is spontaneously broken and it is not if
$E_0=0$.

Suppose that in Eq.~(\ref{one}) one uses instead the periodic boundary
condition (PBC) for all variables. Then the partition function is proportional
to the Witten index $\mathcal{N}_{\text{PBC}}\int_{\text{PBC}}\rmd\mu\,e^{-S}
=\Tr(-1)^Fe^{-\beta H}=\Tr(-1)^F$~\cite{Cecotti:1981fu,Fujikawa:1982nt} and 
$\int_{\text{PBC}}\rmd\mu\,He^{-S}$ is proportional to the $\beta$-derivative of
the Witten index, that is \emph{always\/} zero
\begin{equation}
   \mathcal{N}_{\text{PBC}}\int_{\text{PBC}}\rmd\mu\,He^{-S}
   =\Tr(-1)^FHe^{-\beta H}
   =0.
\label{three}
\end{equation}
This independence of the Witten index on a parameter of the theory~$\beta$ is a
consequence of the supersymmetry algebra~\cite{Witten:1982df}. We also note
that, when the Witten index is non-zero, Eq.~(\ref{three}) is not invariant
under a shift of the origin of the energy $H\to H+c$.
With the periodic boundary condition, we thus have (independently of $\beta$)
\begin{equation}
   \langle H\rangle_{\text{PBC}}\equiv\frac{\int_{\text{PBC}}\rmd\mu\,He^{-S}}
   {\int_{\text{PBC}}\rmd\mu\,e^{-S}}
   =\frac{0}{\Tr(-1)^F}
\label{four}
\end{equation}
and clearly this provides no useful information on the ground-state energy.

In Eq.~(\ref{one}), boundary conditions for realizing the thermal equilibrium
explicitly breaks supersymmetry. In Eq.~(\ref{two}), we then observe how the
effect of the temperature (that is a conjugate variable to the energy) remains
in the zero-temperature limit $\beta\to\infty$. If the effect remains, that is,
$E_0>0$ in Eq.~(\ref{two}), we judge that the spontaneous supersymmetry
breaking occurs. It is interesting to note that this procedure is quite
analogous to a usual way to observe the spontaneous breaking of ordinary
symmetries~\footnote{Note however a crucial difference from ordinary
symmetries; supersymmetry can be broken even with finite spatial
volume~\cite{Witten:1981nf}.}.

There are several issues to be clarified to embody the basic
formula~(\ref{two}) in euclidean lattice formulation. First, the above argument
assumes that a regularization to define the functional integral does not break
supersymmetry. Although lattice regularizations are generally irreconcilable
with supersymmetry, for theories with extended supersymmetry, it is sometimes
possible to set up a lattice regularization that preserves the invariance under
a part of supersymmetry
transformations~\cite{Kaplan:2003uh,Feo:2004kx,Giedt:2006pd,Giedt:2007hz}.
Then, if the spacetime dimension is low enough, one may expect that the
invariance under a full set of supersymmetry transformations is restored in the
continuum limit. In what follows, we assume this sort of lattice
regularization.

Closely related to the above point, we have to properly choose a possible
additive constant in the hamiltonian~$H$. In other words, we have to correctly
choose the origin of the energy. This is of course a crucially important point
to judge the spontaneous supersymmetry breaking from the positivity of~$E_0$. A
natural prescription to define the hamiltonian is to use the supersymmetry
algebra~\footnote{This is precisely the idea of the hamiltonian formulation of
supersymmetric theories (see Ref.~\cite{Beccaria:2004pa} and references cited
therein). From a view point of the gauge invariance, however, the euclidean
lattice formulation appears advantageous.}.
By the following reason, however, this issue of a ``correct'' hamiltonian is
somewhat delicate in the functional integral formulation based on the
lagrangian.

Suppose that the (for simplicity, off-shell) supersymmetry algebra is realized
by the transformation law for variables appearing in the continuum lagrangian.
This implies that there exists a fermionic transformation $Q$ such that
$\{Q,\overline Q\}=2i\partial_0$, where $\overline Q$ is a fermionic
transformation conjugate to $Q$ and $\partial_0$ is the time derivative. One
would then expect from this algebra that the relation
$iQ\overline{\mathcal{Q}}=2H$ holds, where $\overline{\mathcal{Q}}$ is the
Noether charge (in field theory, we use the Noether current instead) associated
with the transformation~$\overline Q$ and $H$ is the hamiltonian obtained from
the lagrangian by the Legendre transformation.

In reality, this relation holds only \emph{up to equations of motion}.
Generally one ends up with $iQ\overline{\mathcal{Q}}/2
=H+(\text{terms being proportional to equations of motion})$. The additional
terms, that would be negligible in classical theory, cannot be neglected in
general within the functional integral because those terms may give rise to
contact terms at a coincident point, i.e., ultraviolet-divergent
constants~\footnote{Interestingly, in models we investigate, the additional
terms cancel with each other and do not contribute to one-point
functions~\cite{Kanamori:2007yx}.}.

However, if a lattice formulation one adopts possesses at least one
exactly-preserved fermionic symmetry, say the above $Q$, it is natural to adopt
$H\equiv iQ\overline{\mathcal{Q}}/2$ as the definition of a hamiltonian.
First, this structure is suggested from the supersymmetry algebra. Second,
this choice has the correct origin of the energy in the sense that it is
consistent with the topological property of the Witten index,
Eq.~(\ref{three}), that is,
\begin{equation}
   \int_{\text{PBC}}\rmd\mu\,He^{-S}
   =\int_{\text{PBC}}\rmd\mu\,Q\left(\frac{i}{2}\overline{\mathcal{Q}}e^{-S}
   \right)
   =0,
\label{five}
\end{equation}
where we have assumed that the lattice action~$S$ and the integration
measure~$\rmd\mu$ are invariant under the $Q$-transformation and the integral
$\int_{\text{PBC}}\rmd\mu\,\overline{\mathcal{Q}}\,e^{-S}$ is finite. As already
noted, when the Witten index is non-zero, this property fixes the origin of the
energy uniquely. For these reasons, we adopt the $Q$-exactness of the
hamiltonian, $H\equiv iQ\overline{\mathcal{Q}}/2$, as a working hypothesis in
what follows.

We examine our idea by applying it to a euclidean lattice formulation of the
supersymmetric quantum mechanics~\cite{Witten:1981nf}. The euclidean lattice
action of this model can be taken
as~\cite{Beccaria:1998vi,Catterall:2000rv,Catterall:2003wd}
\begin{align}
   S&=\sum_{x\in\Lambda}\biggl\{
   \frac{1}{2}\partial\phi(x)\partial\phi(x)
   +\frac{1}{2}(W'(\phi(x)))^2
\nonumber\\
   &\qquad\qquad{}+\overline\psi(x)\left(\partial+W''(\phi(x))\right)\psi(x)
\nonumber\\
   &\qquad\qquad{}-\frac{1}{2}F(x)^2+W'(\phi(x))\partial\phi(x)
   \biggr\},
\label{six}
\end{align}
where $\Lambda=\left\{x\in a\mathbb{Z}\mid0\leq x<\beta\right\}$ ($a$ denotes
the lattice spacing), $\partial$ is the forward difference
$\partial f(x)\equiv f(x+a)-f(x)$ and $\psi(x=\beta)=+\psi(x=0)$ for PBC and
$\psi(x=\beta)=-\psi(x=0)$ for aPBC. This lattice action is invariant under a
lattice counterpart of one of the $\mathcal{N}=2$ supersymmetry
transformations, $Q$, that is defined by $Q\phi(x)=\psi(x)$, $Q\psi(x)=0$,
$Q\overline\psi(x)=F(x)-\partial\phi(x)-W'(\phi(x))$ and
$QF(x)=\partial\psi(x)+W''(\phi(x))\psi(x)$.
The corresponding continuum theory is invariant under also $\overline Q$ and
the supersymmetry algebra reads $Q^2=\overline Q^2=0$ and
$\{Q,\overline Q\}=-2\partial$. Corresponding to this $\overline Q$-symmetry,
we have the Noether charge $\overline{\mathcal{Q}}$. By using a lattice
transcription of this Noether charge
$\overline{\mathcal{Q}}(x)
\equiv-\overline\psi(x)\left(i\partial\phi(x)-iW'(\phi(x))\right)/a$,
we define the hamiltonian operator by
$H(x)\equiv iQ\overline{\mathcal{Q}}(x)/2$.
As elucidated above, this lattice hamiltonian has a correct zero-point energy
in the sense of Eq.~(\ref{five}). In the corresponding continuum
theory~\cite{Witten:1981nf}, it is well-known that supersymmetry is
spontaneously broken if and only if the number of zeros of the function
$W'(\phi)$ is even.

As a definite example, we consider
\begin{equation}
   W(\phi(x))=\frac{1}{2}(am)\phi(x)^2+\frac{1}{3}(am)^{3/2}\lambda\phi(x)^3,
\label{eight}
\end{equation}
where $m$ is a parameter which has the mass dimension~1. Supersymmetry is
dynamically broken in the corresponding continuum theory when $\lambda\neq0$.

Fig.~1 is a result of Monte Carlo simulation for $\lambda=10$. The continuum
limit of the expectation value of the hamiltonian with the anti-periodic
boundary condition, $\lim_{a\to0}\langle H(x)\rangle_{\text{aPBC}}/m$, is plotted
as a function of the physical temporal size of the
system~$\beta m$~\footnote{The result was obtained by an extrapolation to the
continuum $a=0$ by a linear $\chi^2$-fit of data computed at $am=0.1$, $0.05$
and~$0.02$. We used $10^4$ statistically independent configurations for each
set of parameters.}.
\begin{figure}
\includegraphics{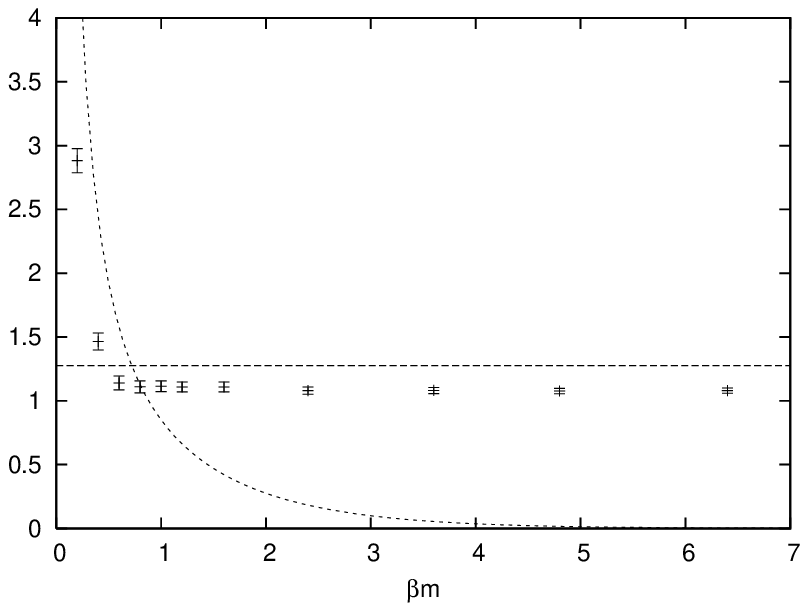}
\caption{$\lim_{a\to0}\langle H(x)\rangle_{\text{aPBC}}/m$ as a function of the
physical temporal size of the system~$\beta m$. The errors are only statistical
ones. The dotted curve is the analytic expression for the $\lambda=0$ case for
which supersymmetry is not broken.}
\end{figure}
For $\beta m\gtrsim1$, we have
$\lim_{a\to0}\langle H(x)\rangle_{\text{aPBC}}/m\simeq1.1$~\footnote{In Fig.~1, we
plotted also the exact ground-state energy $E_0/m=1.27616$ that was obtained by
a numerical diagonalization of the corresponding hamiltonian. The discrepancy
of our Monte Carlo result for $\beta m\gtrsim1$ with this exact result can be
understood by a systematic error associated with the linear extrapolation to
the continuum.}. From this, we infer that supersymmetry is spontaneously
broken and this is indeed the right answer. We also numerically observed that,
in the present system, $\langle H(x)\rangle_{\text{PBC}}$ is not well-defined,
being consistent with $\Tr(-1)^F=0$ in the target theory (recall
Eq.~(\ref{four}))~\cite{Kanamori:2007yx}.

Next, as an example in which supersymmetry is \emph{not\/} spontaneously
broken, we consider
\begin{equation}
   W(\phi(x))=\frac{1}{4}(am)^2\phi(x)^4.
\end{equation}
In this case, we numerically observed that both
$\langle H(x)\rangle_{\text{PBC}}$ and $\langle H(x)\rangle_{\text{aPBC}}$ are
well-defined and, in Fig.~2, we plotted the continuum limit of these quantities
as a function of $\beta m$~\footnote{The result was obtained by an
extrapolation to the continuum $a=0$ by a linear $\chi^2$-fit of data computed
at $am=0.1$ and $0.05$. We used $10^4$ statistically-independent configurations
for each set of parameters.}.
\begin{figure}
\includegraphics{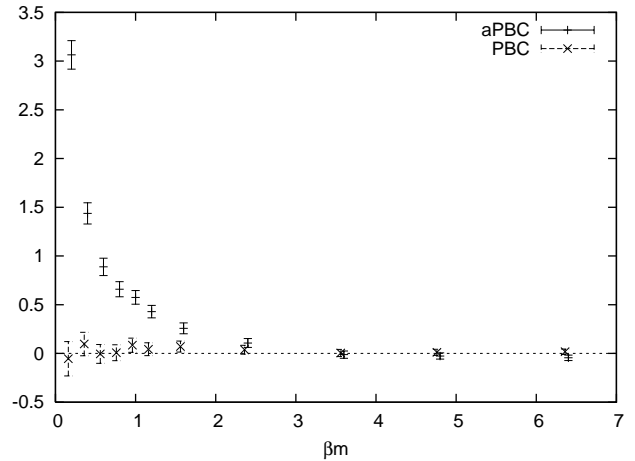}
\caption{$\lim_{a\to0}\langle H(x)\rangle_{\text{aPBC}}/m$ and
$\lim_{a\to0}\langle H(x)\rangle_{\text{PBC}}/m$, as a function of the physical
temporal size of the system~$\beta m$. The errors are only statistical ones.}
\end{figure}
The figure shows that, in this case,
$\lim_{a\to0}\langle H(x)\rangle_{\text{PBC}}$ is consistent with zero for all
temporal sizes (recall Eq.~(\ref{four}); in the target theory $\Tr(-1)^F=1$)
and $\lim_{a\to0}\langle H(x)\rangle_{\text{aPBC}}$ approaches zero as the
temporal size of the system is increased. From Eq.~(\ref{two}), we conclude
that $E_0=0$ within the error and supersymmetry is not broken.

Having observed that our method works perfectly in the supersymmetric quantum
mechanics, we now study the two-dimensional $\mathcal{N}=(2,2)$ super
Yang-Mills theory by using a lattice formulation proposed in
Ref.~\cite{Sugino:2004qd}. (See also Ref.~\cite{Sugino:2003yb}.) For this
seemingly simple supersymmetric system, the value of the Witten index and
whether supersymmetry is spontaneously broken or not are not known, due to
\emph{non-compact\/} flat directions of the classical potential energy. In
fact, the authors of Ref.~\cite{Hori:2006dk} conjectured the dynamical
supersymmetry breaking in this system with the gauge group $\SU(N_c)$.

We numerically studied only the case of the gauge group~$\SU(2)$. The physical
size of our two-dimensional lattice~$\Lambda$ is $\beta\times L$;
$\Lambda=\left\{x\in a\mathbb{Z}^2\mid0\leq x_0<\beta,\,\,0\leq x_1<L\right\}$
and the boundary condition of a generic fermionic field~$\psi$ is set to be,
$\psi(x_0=\beta,x_1)=+\psi(x_0=0,x_1)$ for PBC and
$\psi(x_0=\beta,x_1)=-\psi(x_0=0,x_1)$ for aPBC~\footnote{Space does not permit
to reproduce relevant mathematical expressions and full details of the
simulation. We refer the reader to Ref.~\cite{Kanamori:2007yx} for these and
for a further list of references.}.

The point is that the lattice action and the integration measure of the
lattice formulation of Ref.~\cite{Sugino:2004qd} are manifestly invariant under
a lattice counterpart of a part of the $\mathcal{N}=(2,2)$ supersymmetry
transformations, $Q$. By a similar reasoning as above, a hamiltonian
density~$\mathcal{H}(x)$ is then defined by
$\mathcal{H}(x)\equiv Q\mathcal{J}_0^0(x)/2$, where $\mathcal{J}_0^0(x)$ is a
lattice transcription of the Noether current associated with another fermionic
symmetry of the target continuum theory, $Q_0$. This definition is consistent
with the supersymmetry algebra in the twisted spinor basis,
$\{Q,Q_0\}=2i\partial_0$. From the $Q$-invariance of the lattice action and of
the integration measure, we have
$\int_{\text{PBC}}\rmd\mu\,\mathcal{H}(x)\,e^{-S}=0$ (assuming that the integral
$\int_{\text{PBC}}\rmd\mu\,\mathcal{J}_0^0(x)\,e^{-S}$ is finite) that is
analogous to Eq.~(\ref{five}). Then the ground-state (vacuum) energy
density~$\mathcal{E}_0$ is given by
$\lim_{\beta\to\infty}\lim_{a\to0}\langle\mathcal{H}(x)\rangle_{\text{aPBC}}
=\mathcal{E}_0$. We judge that the dynamical supersymmetry breaking takes place
if $\mathcal{E}_0>0$ and it does not if $\mathcal{E}_0=0$.

Our algorithm and the simulation code, that were developed by using
FermiQCD/MDP~\cite{DiPierro:2000bd,DiPierro:2005qx}, are almost identical to
those of Ref.~\cite{Suzuki:2007jt}.  We use the hybrid Monte Carlo algorithm to
generate configurations in the quenched approximation. The effect of dynamical
fermions is then afterward taken into account by reweighting configurations by
the pfaffian of the Dirac operator (we do not introduce any mass terms of
fermions or bosons that would explicitly break the $Q$-symmetry). Although this
is certainly a brute force method compared to a standard pseudo-fermion
algorithm, its implementation is much simpler and the validity has been
observed for one-point Ward-Takahashi identities~\cite{Suzuki:2007jt}.

The number of statistically-independent configurations we used
is summarized in Table~1, where $N_T$ and $N_S$ are the number of lattice
points for the temporal and the spatial directions, respectively. The physical
size of the spatial direction is fixed to be $Lg=\sqrt{2}$. We used the cold
start and set all scalar fields to be zero at the initial configuration. As
initial thermalization, we discarded the first $10^4$ trajectories and then we
stored configurations at each $10^2$ trajectories (the auto-correlation time
was 10--20 trajectories).
\begin{table*}
\caption{The number of statistically-independent configurations we used.
}
\begin{ruledtabular}
\begin{tabular}{c|c|ccccccc}
\multicolumn{2}{c|}{} & \multicolumn{7}{c}{$N_T/N_S$} \\
\hline
$N_S$ & $ag$ & 0.25 & 0.5 & 1 & 1.5 & 2 & 2.5 & 3 \\
\hline
6     & 0.2357  & ---     & 39,900 & 99,900 & 9,900 &  9,900 & 9,900 & 9,900 \\
8     & 0.1768  & ---     & 39,900 & 99,900 & 9,900 &  9,900 & 9,900 & 9,900 \\
12    & 0.1179 & 39,900 & 69,900 & 69,900 & 9,900 & 9,900 & 9,900 & 9,900 \\
16    & 0.08839 & 39,900 & ---     & ---     & ---    & ---     & --- & --- \\
20    & 0.07071 & 39,900 & ---     & ---     & ---    & ---     & --- & --- \\
\end{tabular}
\end{ruledtabular}
\end{table*}

The result of our Monte Carlo simulation is Fig.~3. The result was obtained,
as Fig.~4, by an extrapolation to the continuum $a=0$ by a linear $\chi^2$-fit.
In Fig.~3, we observe that $\mathcal{E}_0$ is consistent with zero within the
error. We regard this as an indication that supersymmetry is not dynamically
broken in the two-dimensional $\mathcal{N}=(2,2)$ super Yang-Mills theory with
the gauge group~$\SU(2)$. Of course, errors in our present result are still
large and we cannot completely exclude a possibility of the supersymmetry
breaking of $O(1)$ in $\mathcal{E}_0/g^2$. Further reduction of statistical
errors will allow us to conclude whether the scale of the dynamical
supersymmetry breaking is $O(1)$ or not~\footnote{For the present lattice
model, we are at present developing a simulation code with the pseudo-fermion
and the RHMC algorithm. We hope this enable us to reduce the statistical
errors, without increasing the number of configurations substantially.}.
If the above observation of unbroken supersymmetry is true, the expectation
value of the hamiltonian density under the periodic boundary condition
$\langle\mathcal{H}(x)\rangle_{\text{PBC}}$ should be well-defined and vanishes
for all $\beta$. Actually, the real part of the expectation values
$\Real\langle\mathcal{H}(x)\rangle_{\text{PBC}}$ for various $\beta g$ we plotted
in Fig.~3 show that this requirement is met within errors. (The imaginary
part $\Imag\langle\mathcal{H}(x)\rangle_{\text{PBC}}$ is also consistent with
zero and the errors are quite smaller than those for the real part.) This
provides another support for our observation.
\begin{figure}
\includegraphics{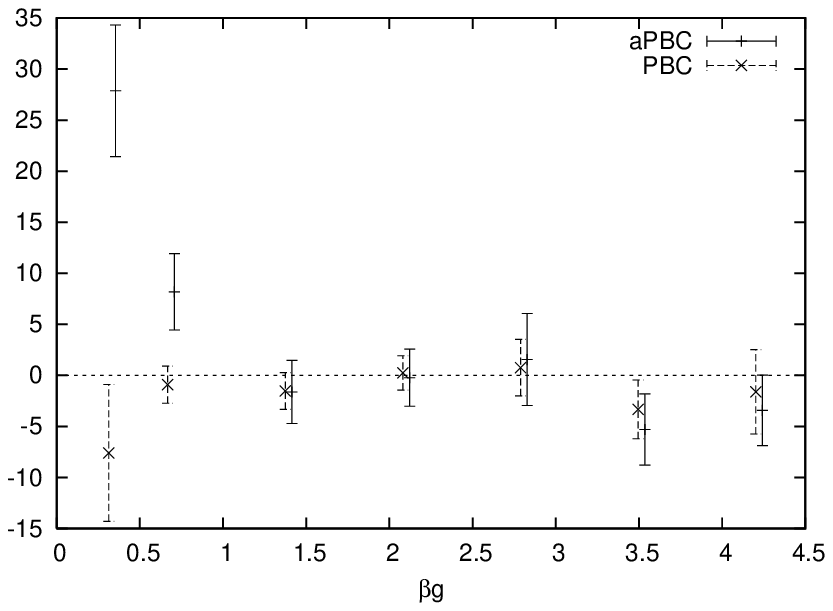}
\caption{$\lim_{a\to0}\Real\langle\mathcal{H}(x)\rangle_{\text{aPBC}}/g^2$ and
$\lim_{a\to0}\Real\langle\mathcal{H}(x)\rangle_{\text{PBC}}/g^2$, as a function of
the physical temporal size of the system~$\beta g$. The errors are only
statistical ones.}
\end{figure}
\begin{figure}
\includegraphics{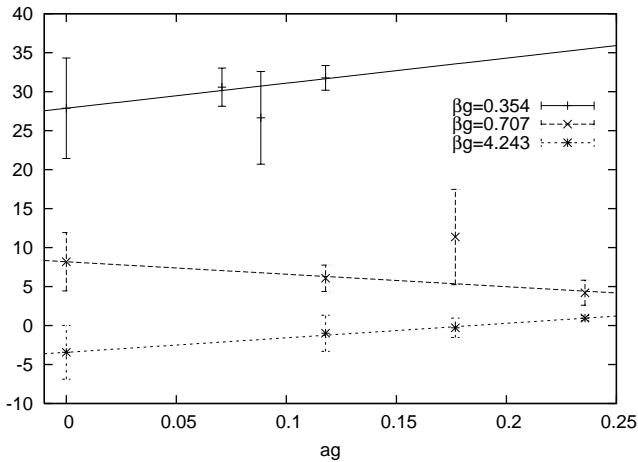}
\caption{Linear $\chi^2$-extrapolation of
$\Real\langle\mathcal{H}(x)\rangle_{\text{aPBC}}/g^2$ to the continuum $a=0$ for
various values of~$\beta g$. The errors are only statistical ones.}
\end{figure}

To our knowledge, this work is the first instance that the dynamical
supersymmetry breaking in a gauge field theory (for which the Witten index is
not known) is investigated by numerical simulation. It should be interesting to
consider applications of the present method to other supersymmetric theories
and gain an insight on a possible supersymmetry breaking that is difficult to
obtain in other ways.

I.K.\ would like to thank Makiko Nio for a useful comment.
F.S.\ would like to thank Kentaro Hori for email correspondence, and the Niels
Bohr Institute for hospitality when this work was in the final stage.
H.S.\ would like to thank Ko Furuta, Masanori Hanada and Tomohisa Takimi for
discussion. The result for the two-dimensional model was obtained by using the
RIKEN Super Combined Cluster (RSCC). I.K.\ is supported by the Special
Postdoctoral Researchers Program at RIKEN. The work of H.S.\ is supported in
part by Grant-in-Aid for Scientific Research, 18540305, and by JSPS and French
Ministry of Foreign Affairs under the Japan-France Integrated Action Program
(SAKURA).

\end{document}